\documentclass{aa}

\usepackage{graphicx}
\usepackage{txfonts}
\newcommand{\hii}{\relax \ifmmode {\mbox H\,{\scshape ii}}\else H\,{\scshape ii}\fi}
\newcommand{\mi}{\relax \ifmmode {\mu{\mbox m}}\else $\mu$m\fi}
\newcommand{\ha}{\relax \ifmmode {\mbox H}\alpha\else H$\alpha$\fi}
\newcommand{\hb}{\relax \ifmmode {\mbox H}\beta\else H$\beta$\fi}

\newcommand{\sii}{\relax \ifmmode {\mbox S\,{\scshape ii}}\else S\,{\scshape ii}\fi}
\newcommand{\siii}{\relax \ifmmode {\mbox S\,{\scshape iii}}\else S\,{\scshape iii}\fi}
\newcommand{\siv}{\relax \ifmmode {\mbox S\,{\scshape iv}}\else S\,{\scshape iv}\fi}
\newcommand{\oiv}{\relax \ifmmode {\mbox O\,{\scshape iv}}\else O\,{\scshape iv}\fi}
\newcommand{\nii}{\relax \ifmmode {\mbox N\,{\scshape ii}}\else N\,{\scshape ii}\fi}
\newcommand{\niii}{\relax \ifmmode {\mbox N\,{\scshape iii}}\else N\,{\scshape iii}\fi}
\newcommand{\oi}{\relax \ifmmode {\mbox O\,{\scshape i}}\else O\,{\scshape i}\fi}
\newcommand{\oii}{\relax \ifmmode {\mbox O\,{\scshape ii}}\else O\,{\scshape ii}\fi}
\newcommand{\hei}{\relax \ifmmode {\mbox He\,{\scshape i}}\else He\,{\scshape i}\fi}
\newcommand{\heii}{\relax \ifmmode {\mbox He\,{\scshape ii}}\else He\,{\scshape ii}\fi}
\newcommand{\oiii}{\relax \ifmmode {\mbox O\,{\scshape iii}}\else O\,{\scshape iii}\fi}
\newcommand{\ariii}{\relax \ifmmode {\mbox Ar\,{\scshape iii}}\else Ar\,{\scshape iii}\fi}
\newcommand{\arv}{\relax \ifmmode {\mbox Ar\,{\scshape v}}\else Ar\,{\scshape v}\fi}
\newcommand{\arii}{\relax \ifmmode {\mbox Ar\,{\scshape ii}}\else Ar\,{\scshape ii}\fi}
\newcommand{\ariv}{\relax \ifmmode {\mbox Ar\,{\scshape iv}}\else Ar\,{\scshape iv}\fi}
\newcommand{\neii}{\relax \ifmmode {\mbox Ne\,{\scshape ii}}\else Ne\,{\scshape ii}\fi}
\newcommand{\neiii}{\relax \ifmmode {\mbox Ne\,{\scshape iii}}\else Ne\,{\scshape iii}\fi}
\newcommand{\nev}{\relax \ifmmode {\mbox Ne\,{\scshape v}}\else Ne\,{\scshape v}\fi}

\newcommand{\rdostres}{\relax \ifmmode {\,\mbox{R}}_{\rm 23}\else \,\mbox{R}$_{\rm 23}$\fi}

\newcommand{\ciii}{\relax \ifmmode {\mbox O\,{\scshape iii}}\else C\,{\scshape iii}\fi}
\newcommand{\civ}{\relax \ifmmode {\mbox O\,{\scshape iii}}\else C\,{\scshape iv}\fi}

\newcommand{\gsim}{\hbox{\rlap{\lower.55ex\hbox{$\sim$}} \kern-.3em
\raise.4ex \hbox{$>$}}}
\newcommand{\lsim}{\hbox{\rlap{\lower.55ex\hbox{$\sim$}} \kern-.3em
\raise.4ex \hbox{$<$}}}

\begin{document}

   \title{Exploring the hardness of the ionising radiation with the infrared softness diagram}
\subtitle{I. Similar effective temperature scales for starbursts and (ultra)luminous infrared galaxies}
\titlerunning{The infrared softness diagram in SFGs and (U)LIRGs}

\author{
        E.~P\'{e}rez-Montero\inst{\ref{IAA}}  \and
        J.~A.~Fern\'andez-Ontiveros\inst{\ref{CEFCA}} \and
        B.~P\'{e}rez-D\'{i}az\inst{\ref{IAA}}  \and
        J.~M. V\'{i}lchez\inst{\ref{IAA}}  \and
        N.~Kumari\inst{\ref{STScI}} \and
        R.~Amor\'{i}n\inst{\ref{CEFCA},\ref{ARAID}} 
}

\institute{
Instituto de Astrof\'{i}sica de Andaluc\'{i}a (CSIC), Apartado 3004, 18080 Granada, Spain \label{IAA}
\and
Centro de Estudios de Física del Cosmos de Aragón (CEFCA), Unidad Asociada al CSIC, Plaza San Juan 1, E--44001 Teruel, Spain\label{CEFCA}
\and
ESA for AURA, STScI, Baltimore, MD 21218, USA\label{STScI}
\and
ARAID Foundation, Avda de Ranillas, 1-D, planta 2ª, oficina b. 50018 Zaragoza, Spain\label{ARAID}
}

   \date{Received XXX; accepted YYY}


  \abstract
   {}
 {We explored the {softness parameter} in the infrared, $\eta\prime_{IR}$, whose main purpose is the characterisation of the hardness of the incident ionising radiation in emission-line nebulae. This parameter is obtained from the combination of mid-infrared wavelength range transitions corresponding to consecutive ionisation stages in  star-forming regions.}
{We compiled  observational data from a sample of star-forming galaxies (SFGs), including luminous and ultraluminous infrared galaxies (LIRGs and ULIRGs), to study the  $\eta\prime_{IR}$ and its equivalent expression in two dimensions, the {\em softness diagram}. We compared them  with predictions from photoionisation models to
determine the shape of the ionising continuum
energy distribution in each case.
We also used the measured  emission-line ratios as   input for {\sc HCmistry-Teff-IR}, a code that performs a  Bayesian-like comparison with photoionisation model predictions in order to quantify the equivalent effective temperature ($T_*$) and the ionisation parameter.}
        {We found similar average values within the errors 
of $\eta\prime_{IR}$ in (U)LIRGs (-0.57)  in the rest of the SFGs (-0.51), which could be interpreted as indicative of a similar incident radiation field. 
This result is confirmed from the analysis using {\sc HCm-Teff-IR}, 
which simultaneously points to a slightly lower, although similar within the errors, $T_*$ scale for (U)LIRGs, even when a higher dust-to-gas mass ratio is considered in the models for these objects. These derived $T_*$ values are compatible with  the ionisation from massive stars, without any need of harder ionising sources, both for (U)LIRGs and the rest of the SFGs. 
However, the derived $T_*$ in (U)LIRGs do not show any correlation with metallicity. This could be interpreted as a sign that their similar average $T_*$ values are due to the attenuation of the energetic incident flux from massive stars by the heated dust mixed with the gas. This is supported by the known very large amounts of small grains associated with the very high star formation rates measured in galaxies of  this type.}
  {}


   \maketitle
%

\section{Introduction}

Emission-line nebulae, including those ionised by episodes of massive star formation and/or by the presence of active galactic nuclei (AGN), can provide us with very valuable information about the several physical processes that take place in galaxies at different scales and distances. This information can  help to disentangle the mechanisms governing their formation and evolution.

Bright optical emission-line intensities, of both recombination and collisional nature, can be used to determine the physical and chemical properties of the ionised gas of the objects where they are measured, given the easier access to this spectral regime from ground-based telescopes. Moreover, the emission lines in the infrared (IR) spectral regime can also supply   very valuable information that
can enhance and complement that obtained from the optical regime.

Among the different advantages entailed by the study of the IR lines is that  they can trace larger optical depths in the ionised nebulae as extinction is attenuated in this regime. In addition, as most of the observed emission lines in the near- and mid-IR ranges have a fine-structure nature close to the ground state, their fluxes  present on average much lower dependence on the electron temperature of the gas   compared to optical collisionally-excited emission lines (CELs). This dependence of the fluxes of optical CELs on temperature is what makes  the determination of chemical abundances so uncertain, as the cooling rate of the gas is dominated by its overall metal content. Finally, we can find in the IR range high-excitation emission lines not measurable in the optical, which can also lead to deriving the ionic abundances of unseen ionisation stages of certain chemical elements, or to better quantifying the excitation stage of the gas in highly ionised environments, such as in the surroundings of an AGN, {using lines such as [\nev] at $\lambda$ 14.3 $\mu$m and $\lambda$ at 24.3 $\mu$m,  or [\oiv] at $\lambda$ 25.9 $\mu$m (e.g. \citealt{Satyapal21})}.

In addition to chemical abundances or excitation, the measured emission-line fluxes in any spectral regime can be used to estimate other properties such as the hardness of the incident radiation field. This method can provide accurate answers regarding the nature of the very young ionising stellar population  compared to  other techniques based on the study of the stellar continuum. In this way the {\em softness parameter}, denoted as $\eta$, was defined by \cite{vp88} to quantify the shape of the incident spectral energy distribution (SED), relating different ratios of ionic abundances of consecutive ionic stages, corresponding to different ionisation potentials. The form of this parameter, originally defined for the optical range, is
\begin{equation}
\eta = \frac{{\rm O}^+/{\rm O}^{2+}}{{\rm S}^+/{\rm S}^{2+}}
.\end{equation}

\begin{figure*}
   \centering
\includegraphics[width=0.9\textwidth,clip=]{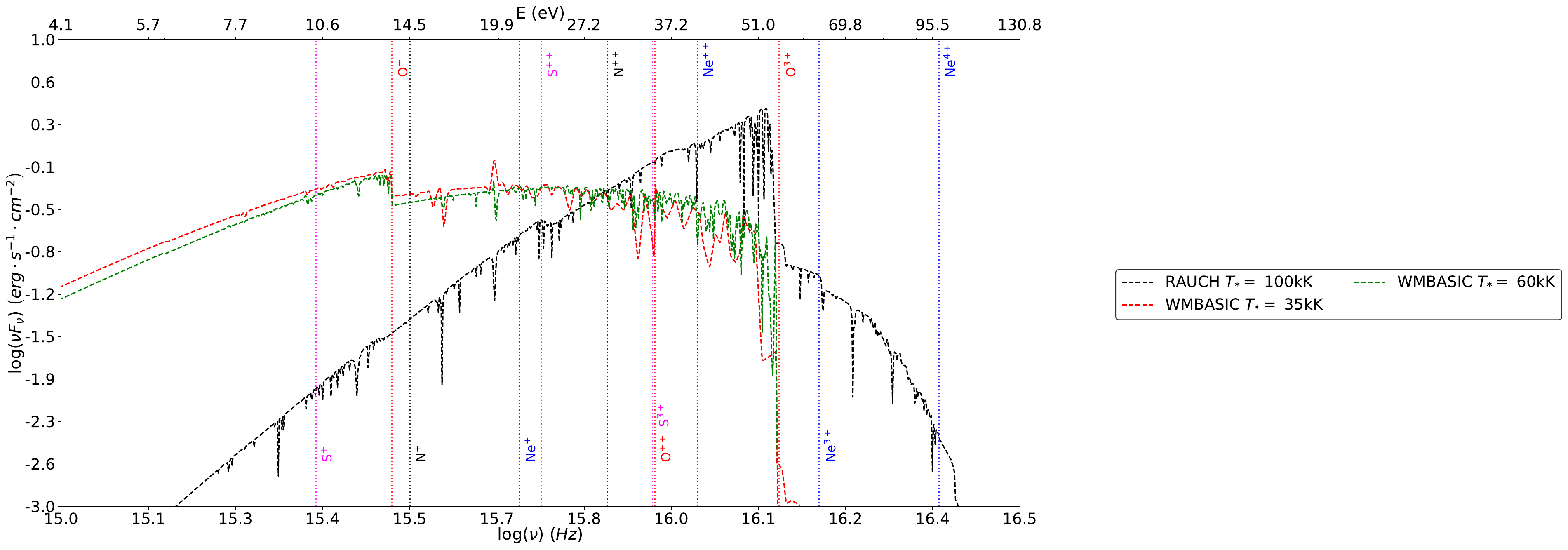}

\caption{Spectral energy distributions of some of the sources considered by   photoionisation models for star-forming galaxies with the energies of the ionisation potentials of the ions whose emission lines are used in our calculations in the optical and IR ranges.}

              \label{SEDs}%
    \end{figure*}

{This expression presents almost no dependence on the ionisation parameter (i.e. the ratio of the number of ionising photons to the density of particles),
and can trace the hardness of the incident radiation field as the involved ionisation potentials of the ions in the numerator are higher than those in the denominator, as can be seen in Figure \ref{SEDs}.}
Hence, lower values of $\eta$ can be taken as a sign of a harder SED. Alternatively, an equivalent form of this parameter can be defined as a function of the corresponding emission-line fluxes (i.e. [\oii], [\oiii], [\sii], [\siii] in the optical and near-IR), denoted as $\eta\prime$, but in this case the parameter presents an additional dependence on metallicity, {due to the collisional nature of the involved emission lines} (e.g. \citealt{kumari21}).

In addition, several different forms of $\eta\prime$ can also be defined in the mid-IR regime based on the Ar, Ne, or S lines \citep{martin02,morisset04}.
In particular, $\eta\prime_{IR}$  can be defined  using Ne and S emission lines in the following way:
\begin{equation}
\eta\prime_{IR} = \frac{{\rm [NeII]} \lambda 12.8 \mu{\rm m}/{\rm [NeIII]} \lambda 15.6 \mu{\rm m}}{{\rm [SIII]} \lambda 18.7 \mu{\rm m}/{\rm [SIV]} \lambda 10.5 \mu{\rm m}}
.\end{equation}This form of the parameter is especially convenient for local nearby galaxies as all their involved  lines are covered by {\em Spitzer} or {\em JWST}.

Moreover, the study of $\eta\prime_{IR}$ presents additional advantages  compared to other similarly defined expressions using only optical emission lines. As studied by \cite{pmv09}, observations are well covered by sequences of photoionisation models with massive single stars
for different values of the equivalent effective temperature ($T_*$) when these are analysed in the so-called {\em softness diagram} (i.e. involving in each axis the corresponding emission-line ratios). This results seems to be contradictory with the equivalent diagram based on optical emission lines, which suggests that additional sources of ionisation, akin to having a radiation field of greater intensity than if it were solely attributed to massive stars, may be necessary to reproduce the position of some star-forming regions (e.g. HOLMES; \citealt{kumari21}). However, as analysed in \cite{pm23}, the use of some low-excitation lines, such as [\sii] in this optical diagram can be contaminated by diffuse ionised gas (DIG) in regions observed using  low spatial resolution, which could be the cause of the very low observed $\eta\prime$ values.  
This result could thus reinforce the idea that diagrams based on high-excitation lines could be more accurate as they probe regions closer to the ionising stars.

Nevertheless, using the softness parameter to analyse the hardness of the incident radiation must take into account that additional  dependences on other properties of the ionised gas (e.g. metallicity, excitation, or the nature of the source) exist.
For this reason, \cite{hcm-teff} introduced the code {\sc HII-CHI-mistry-Teff} (hereafter {\sc HCm-Teff}), which decomposes these dependences using a Bayesian-like comparison between the fluxes of the emission lines involved in the softness diagram, with the predictions from large grids of photoionisation models dependent on different values of $T_*$, the ionisation parameter $U$, and the metallicity. In this work we describe the adaptation of this code for the IR emission lines (i.e. {\sc HII-CHI-mistry-Teff-IR}, hereafter {\sc HCm-Teff-IR}), which is particularly useful for investigating the nature of the ionising stellar population in star-forming galaxies, especially those obscured by large amounts of dust, such as (ultra)luminous infrared galaxies [(U)LIRGs].

These galaxies offer a valuable opportunity to study extreme environments and to unravel the underlying processes driving their energetic output, including very intense bursts of star formation triggered in most cases by mergers or interactions (e.g. \citealt{sm96}) leading to their extreme IR luminosities. Thus, understanding the nature of the stellar populations within these galaxies is fundamental to comprehending their ionising spectra and the mechanisms responsible for their extraordinary luminosities.

Although hugely obscured by the great amount of dust present in these objects, the synthetic spectra fitting study of the optical spectrum in (U)LIRGs, \citep{rodzau10,pereira15} shows that the ionising stellar population of their nuclei has a  young mean stellar age lower  than 100 Myr, and more intense than previous episodes of star formation, likely a consequence of the initial interactions preceding the final merger. The spatially resolved study of these galaxies also reveals an important contribution of DIG to the low-excitation lines from beyond the nuclear regions \citep{ah10}.
This DIG component could be featured by smaller dust optical depths and heated by stars older than 100 Myr \citep{dacunha10}. 

The study of the hardness of the ionising incident radiation using IR emission lines (less attenuated by dust absorption) and high-excitation transitions (less affected by DIG contamination) offers a unique opportunity to probe the nature of the very obscured young stellar populations in these objects. Therefore, we take advantage of the new methodology based on IR lines described in this work to perform a study of the incident radiation in a sample of (U)LIRGs dominated by star formation, to be compared with the rest of the star-forming galaxies (SFGs).

The paper is organised as follows. In Section 2 we describe the observational samples of compiled near- and mid-IR emission lines in the sample of star-forming galaxies and (U)LIRGs. In Section 3 we describe the grids of photoionisation models used to derive emission-line fluxes in the IR to construct the different versions of the softness diagram, and the code {\sc HCm-Teff} for the IR to extract the corresponding values dominating this diagram for SFG. In Section 4 we present our results from the application of this code to the different samples of objects. Finally, in Section 5 we summarise our results and give our conclusions.

%

\section{Data samples}

We collected emission-line fluxes in the mid- and far-IR spectral ranges for star-forming galaxies and (U)LIRGs dominated by star formation from two sources. The first  is \cite{jafo21}, who gathered observations in the mid- to far-IR range acquired with the InfraRed Spectrograph (IRS; \citealt{houck2004}) on board \textit{Spitzer} and with the Photodetector Array Camera and Spectrometer (PACS; \citealt{poglitsch2010}) on board \textit{Herschel}. A few detections of the [\ion{O}{iii}]$_{\rm 52 \mu m}$ and [\ion{N}{iii}]$_{\rm 57 \mu m}$ far-IR lines obtained with FIFI-LS \citep{fischer2018} on board the \textit{SOFIA airborne observatory} \citep{temi2018} were also taken from \citet{peng2021}. This sample collected by \cite{jafo21} includes 28 dwarf galaxies with subsolar metallicities \citep{madden2013,cormier2015}, 19 active star-forming galaxies with near solar metallicities \citep{jafo2016}, and 9 low-redshift (U)LIRGs dominated by the star formation component \citep{pereira-santaella2017,jafo2016}. The AGN contribution in these galaxies is lower than $10\%$, according to their [\ion{Ne}{v}]$_{\rm 14.3 \mu m}$ / [\ion{Ne}{ii}]$_{\rm 12.8 \mu m}$ measured ratios \citep{pereira-santaella2017}.

The second source includes objects from the IDEOS catalogue \citep{HernanCaballero_2016, Spoon_2022} IR database, which includes the measurements of 77 fitted mid-IR spectral features in the 5.4-36 $\mu$m range for all galaxies observed with \textit{Spitzer} (a total of 3335 galaxies; \citealt{Spoon_2022}). From this initial sample, 153 star-forming objects (P\'erez-D\'\i az et al. submitted) and 66 (U)LIRGs \citep{pd23a} were selected by omitting duplicate observations of the same object and entries corresponding to galaxies presented in the first catalogue, where the higher spectral resolution observation was used, 
and after verifying that star formation dominates the ionisation budget based on the strength of polycyclic aromatic hydrocarbon (PAH) features (EW(PAH6.2 $\mu$m) $>$ 0.06 $\mu$m) and the small contribution from certain highly ionised species ([\nev] $\lambda$ 14$\mu$m/[\neii] $\lambda$12$\mu$m $<$ 0.15 and [\oiv] $\lambda$ 26$\mu$m/[\neii] $\lambda$ 12$\mu$m $<$ 0.4).
Our final sample then comprises 228 star-forming galaxies and 75 (U)LIRGs from both sources.

{We collected total oxygen abundances for all objects in this selected sample  as derived by P\'erez-D\'\i az et al. (submitted). These abundances}
were computed  using version 3.2  of the code {\sc HII-CHI-mistry-IR} (hereafter {\sc HCm-IR}; \citealt{jafo21,borja22}).\footnote{All versions of the code {\sc HCm} can be retrieved in the web page \url{http://home.iaa.es/~epm/HII-CHI-mistry.html}}
Unlike previous versions of {\sc HCm-IR}, version 3.2  does not consider S emission lines to derive oxygen abundances as these lines are used in the new version for the derivation of sulphur abundances.
This improvement of the code is described and discussed in P\'erez-D\'\i az et al. (submitted), where it is   checked that this change does not imply results in the derived oxygen abundances deviating beyond the obtained errors  compared to the values derived using previous versions of the code, {such as the values published by \cite{jafo21}.}

The code   performs a Bayesian-like calculation using certain observed emission-line ratios sensitive to the metallicity and the excitation compared with the predictions from a large grid of photoionisation models. 
The code gives as a result the means of the nitrogen-to-oxygen abundance ratio (N/O), the total oxygen abundance [12+log(O/H)], and the ionisation parameter (log $U$), calculated as the averages of the resulting $\chi^2$-weighted distributions, with the corresponding uncertainties calculated as the quadratic sum of the standard deviation of the same distribution and the dispersion associated to a Monte Carlo simulation considering the observational reported errors.

For this work we compared the compiled observations with the predictions from the photoionisation code {\sc Cloudy} v17.1 \citep{cloudy}, assuming {\sc Popstar} \citep{popstar} synthesis model atmospheres. We considered as templates for the grids of models those defined in \cite{hcm14} to establish empirical relations between O/H, N/O, and $U$ used by the code in the absence of any auroral emission line.
The resulting metallicity distribution was later used to constrain the final $T_*$ and $U$ values in the softness diagram, as described in the sections below.

\section{The {\sc HCm-Teff-IR} code}

\begin{figure}
   \centering
\includegraphics[width=8cm,clip=]{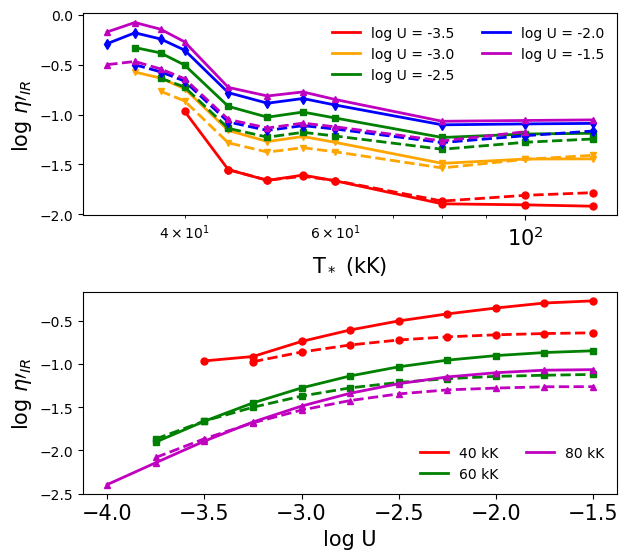}

\caption{Dependence of the $\eta\prime_{IR}$ parameter as a function of $T_*$ (upper panel) and log $U$ (lower panel) according to the photoionisation models calculated using WM-Basic (30-60 kK) and \cite{rauch} (80-120 kK) SEDs.
The solid (dashed) lines represent models calculated for 12+log(O/H) = 8.9 (8.0).}

              \label{eta-IR}%
    \end{figure}

\subsection{Description of the photoionisation models}

In order to obtain an estimate of the hardness of the incident radiation field, we calculated several different grids of photoionisation models whose predicted IR line intensities were later compared by the code with the compiled IR data described in the previous section. The quantification of the hardness of the incident field of radiation was performed later   using as a parameter the equivalent effective temperature ($T_*$), so  in the models we used SEDs from single stars or black bodies, {whose $T_*$ can be more easily parametrised than the synthetic massive stellar clusters used for the derivation of chemical abundances, which does not imply deviations larger than the obtained uncertainties in the results, as discussed in  \citep{hcm-teff}.} 

The   photoionisation  models used are the same as described in \cite{hcm-teff} adopting SEDs for massive single-stars from WM-Basic \citep{wmbasic} in the range of $T_*$ between 30 kK and 60 kK. In addition, we incorporated models from \cite{pm23} using central stars of planetary nebulae (PNe) SEDs from \cite{rauch} in the range of $T_*$ from 80 kK to 120 kK.

The models cover a range in total oxygen abundance from 12+log(O/H) = 7.1 to 8.9, and log $U$ from -4.0 to -1.5. 
All other chemical abundances were scaled to  solar proportions, as described by \cite{asplund09}, with the exception of N, which follows the empirical relation derived by \cite{hcm14} (i.e. a constant N/O ratio due to primary N production for low metallicity, and an increasing N/O due to secondary N production for higher metallicity). 
{A particle density of 100 cm$^{-3}$ and a filling factor of 0.1 were assumed in all models.
The dust-to-gas mass ratio was assumed to have standard galactic proportions ($d/g_{MW}$ = 7.48 $\cdot$ 10$^{-3}$).
In addition, to consider possible higher dust proportions in the case of (U)LIRGs,  more grids were computed assuming a $d/g$ = $2\cdot d/g_{MW}$. However, as in other works describing models for this dust-bounded objects (e.g. \citealt{abel09}) we did not explore other grain size distributions or depletion factors, so in our models we only refer  to gas-phase abundances.} All models were calculated assuming a plane-parallel and a spherical geometry to inspect the impact of this in the results. In addition, the stopping criterion of all models is that the fraction of free electrons is not lower than 98\%.  

The resulting emission-line fluxes were later compiled in tables in order to be compared with the observed values using the code {\sc HCm-Teff-IR}, adopting the same procedure  described in \cite{hcm-teff} for the case of the optical lines.

\subsection{Description of the code }

The use of the softness parameter based on IR lines, as defined in equation 2, or using alternative ratios of emission-line fluxes corresponding to consecutive states of ionisation, can be used as a proxy for the hardness of the incident radiation. The $\eta\prime_{IR}$ parameter based on [\neii], [\neiii], [\siii], and [\siv], can trace the shape of the incident SED in an energetic range quite similar to that covered by the parameter based on optical lines. In Figure \ref{SEDs} it can be seen that the  ionisation potentials corresponding to the emission lines involved in both the optical and IR softness parameters trace similar energetic ranges, although slightly higher as S$^+$ is replaced with S$^{3+}$. Nonetheless, this replacement is fundamental to greatly reducing the influence of DIG emission on the new parameter based on IR lines, given the known enhancement of 
[\sii] emission (relative to \ha) in the DIG with respect to the
         \hii\ regions \citep{reynolds85, dm94, galarza99}. In this way, as shown by models in Figure \ref{eta-IR}, the $\eta\prime_{IR}$ parameter can be used to estimate the hardness of the radiation field, which  decreases for increasing values of $T_*$, as traced by WM-Basic SEDs in the 30-60 kK range and by \cite{rauch} in the 80-120 kK range.

However, as discussed in \cite{hcm-teff} for the optical version of the equivalent parameter based on [\oii], [\oiii], [\sii], and [\siii], $\eta\prime$ presents additional dependences on other functional parameters that can lead to incorrect interpretations of objects with different $\eta\prime$ values, but also with very different metallicities or excitations. Similarly, as   can   be seen in the same figure, $\eta\prime_{IR}$ also varies significantly for sequences of the same $T_*$ with different excitation, as $\eta\prime_{IR}$ increases with log $U$. In addition, the parameter is also sensitive to the assumed metallicity as it is lower for lower O/H, although this dependence is much lower than in the case of the equivalent optical parameter and is only clear for high values of $U$.

In this scenario, a code like {\sc HCm-Teff} for the IR can find a more accurate solution for $T_*$ by performing a Bayesian comparison between the observed emission-line ratios and the grid of photoionisation models. When the parameters of the models that will be used have been chosen (i.e. SED, geometry), and if metallicity is given as an input expressed in the form of total oxygen abundance, the code firstly interpolates for each entry the tables of the emission-line fluxes predicted by the models for the corresponding metallicity. Otherwise, if the metallicity is not specified, the code performs the calculation over all the possible values in the tables.

Then the code creates a distribution of $T_*$ and log $U$ weighted by the corresponding $\chi^2$ obtained as the sum of the quadratic difference of the observed and the predicted emission-line ratios that can be calculated with the emission-line fluxes given as input. The code gives as solutions for $T_*$ and log $U$ the mean of the obtained weighted distributions, and as errors the quadratic sum of the corresponding standard deviation of the distributions and the dispersion of the solutions found after a Monte Carlo iteration, using as input values the nominal fluxes perturbed with the observational errors.

The code takes as input the fluxes in arbitrary units of the emission lines in the following ratios: {[\arii]} (7.0 $\mu$m) and [\ariii] (9.0 $\mu$m and/or 21.8 $\mu$m); {[\neii]} (12.8 $\mu$m) and [\neiii] (15.6 $\mu$m); {[\siii]} (18.7 $\mu$m and/or 33.5 $\mu$m) and [\siv] (10.5 $\mu$m); {[\oiii]} (52 $\mu$m and/or 88 $\mu$m) and [\oiv] (25.9 $\mu$m);
and, {[\niii]} (57 $\mu$m) and [\nii] (122 $\mu$m and/or 205 $\mu$m). The code only considers an emission-line flux for the calculation when it can be used in combination with the corresponding emission-line flux in the consecutive ionisation stage.  In our case no reddening correction was applied to the IR line fluxes given the minor effect of dust extinction in this spectral range.

\begin{figure}
   \centering
\includegraphics[width=8cm,clip=]{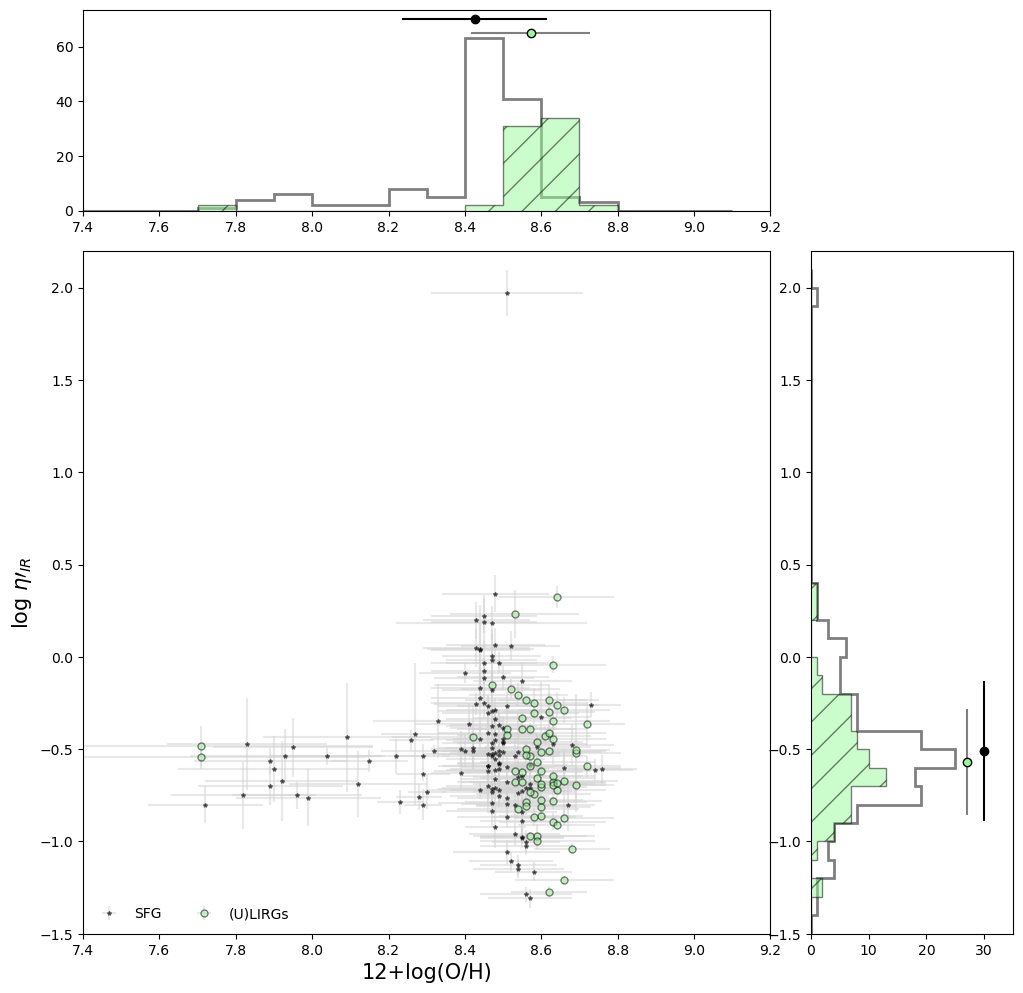}

\caption{{Relation for the studied sample between the derived total oxygen abundance and log $\eta\prime_{IR}$, as defined in equation 2. The black stars represent SFGs and green circles (U)LIRGs. The top and right subpanels show the corresponding distributions and the mean values of each variable.   The black empty histograms represent SFGs, while the hatched green histograms represent (U)LIRGs.}}

    \label{eta-OH}%
    \end{figure}

\section{Results and discussion}

\subsection{The IR softness parameter}

We investigated how the combined use of emission-line ratios of consecutive states observable in the mid-IR can be used to determine the hardness of the incident radiation in emission-line objects ionised by massive episodes of star formation. In particular, we paid special attention to possible   differences in   the subsample of (U)LIRGs  compared to the rest of the star-forming galaxies.

As a first step we studied the behaviour of the $\eta\prime_{IR}$ parameter, as defined in equation 2, for
the samples of galaxies described in section 2 in order to find possible evidence  for a different hardness of the incident ionising radiation in (U)LIRGs   compared with the rest of  the SFGs.

We show in Figure \ref{eta-OH} the obtained log $\eta\prime_{IR}$ values as a function of the derived oxygen metallicity.
The distribution of this latter derived in each sample using IR lines with {\sc HCm-IR} \citep{jafo21} is also shown in the uppeer panel of the same figure, along with their mean values, {while the distributions and means for log $\eta\prime_{IR}$ are shown in the right panel. }  

Regarding the softness parameter, both samples show a large dispersion, but a clear offset between the two samples cannot be appreciated. This can also be easily seen by comparing the very similar mean values of log($\eta\prime_{IR}$) for the subsamples of SFGs (-0.51)  and (U)LIRGs (-0.57), with a difference that is  much lower than the error bars corresponding to the standard deviation of the distributions (i.e. 0.38 for SFGs and 0.28 for (U)LIRGs) overlapping in a long range.

Therefore, although the value of log $\eta\prime_{IR}$ measured in the ULIRG sample is slightly lower than for the rest of the SFGs, which  could be in principle interpreted as a harder ionising field of radiation, the difference between their means is not larger  than the associated dispersions, and
it does not thus seem to be significant,  at least from the unique interpretation of this parameter. 
Moreover, this very small difference between the obtained $\eta\prime_{IR}$ values found for (U)LIRGs  compared with the other SFGs seems to be consistent with
the apparent lack of a correlation with metallicity, and with that the average oxygen abundances found in each sample are similar considering the dispersion (i.e. 12+log(O/H) = 8.48 for SFGs and 8.58 for (U)LIRGs).

However, as discussed in Section 3, the softness parameter presents additional dependences that must be considered. 
For instance, as shown in Figure 2, $\eta\prime_{IR}$ tends to be lower for lower metallicity, although this is only noticeable for high excitation. It is thus necessary to evaluate if the obtained higher average oxygen abundance measured in (U)LIRGs could imply that the average incident field of radiation in this objects could be even harder than the trend suggested by the  use of $\eta\prime_{IR}$ alone, not significant by itself considering the uncertainties.

Therefore, in order to correctly interpret whether the  observed similarity of $\eta\prime_{IR}$ between (U)LIRGs and the rest of the SFGs is significant, it is necessary to compare this observational trend with the predictions from large grids of photoionisation models that take into account the differences in metallicity and excitation.

\begin{figure*}
   \centering
\includegraphics[width=0.4\textwidth,clip=]{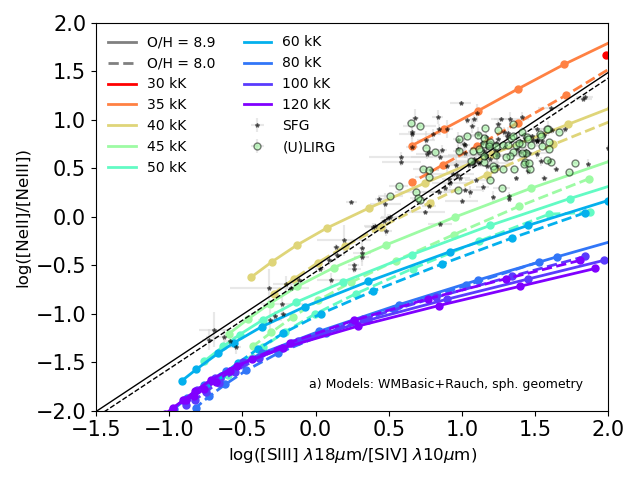}
\includegraphics[width=0.4\textwidth,clip=]{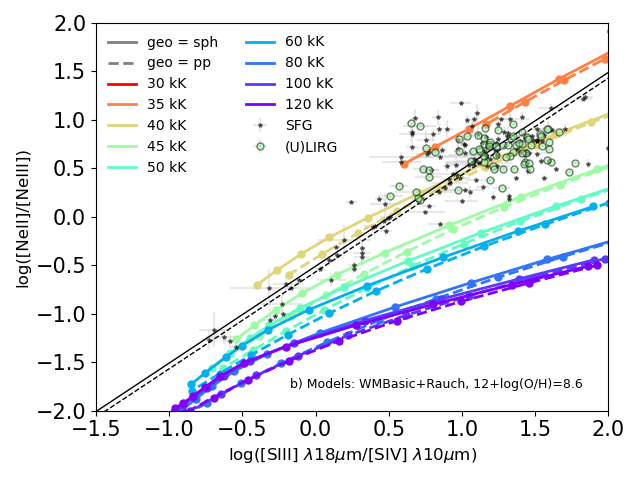}
\includegraphics[width=0.4\textwidth,clip=]{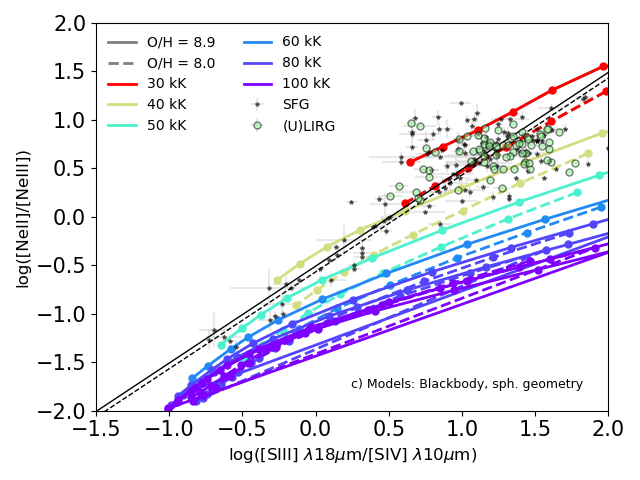}
\includegraphics[width=0.4\textwidth,clip=]{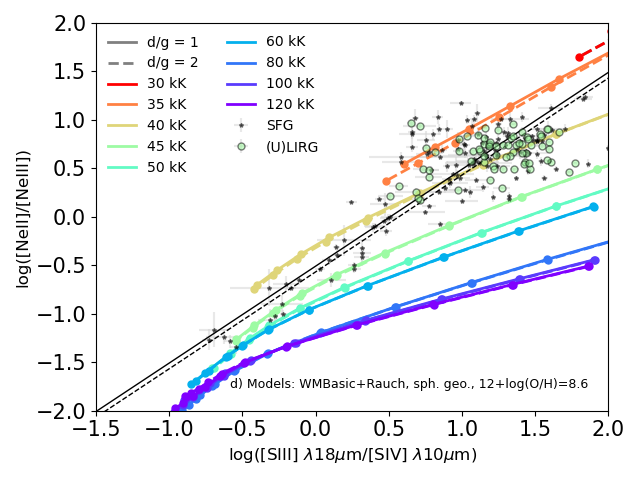}

\caption{Infrared softness diagram based on the emission-line ratios [\neii] $\lambda$ 12.8 $\mu$m/[\neiii] $\lambda$ 15.6 $\mu$m and [\siii] $\lambda$ 18.7 $\mu$m /[\siv] $\lambda$ 10.5 $\mu$m for the sample of objects described in the text: star-forming galaxies (black stars) and (U)LIRGs (green circles). The black solid and dashed lines represent the mean values of the $\eta\prime_{IR}$ parameter for SFGs and (U)LIRGs, respectively. Shown in  each panel are the sequences of models  for different values of $T_*$ {with different conditions to be compared with the data, including: a)   Models calculated with  WM-Basic \citep{wmbasic} massive stars (from 30 to 60 kK) and \cite{rauch} central stars of PNe (from 80 to 120 kK)  assuming a spherical geometry. The solid lines represent models with 12+log(O/H) = 8.9, while the dashed lines represent a value of 8.0. b)  WM-Basic and Rauch models with 12+log(O/H) = 8.6 with spherical (solid line) and plane-parallel (dashed) geometry. c)  Models calculated assuming a black-body incident continuum from 30 kK up to 100 kK assuming spherical geometry with 12+log(O/H) = 8.9 (solid line) and 8.0 (dashed line). d)  Models using WM-Basic and Rauch stellar atmospheres with spherical geometry and 12+log(O/H) = 8.6, and a standard value for the dust-to-gas mass ratio (solid line), and with a doubled dust content (dashed line).} }

    \label{eta-sf}%
    \end{figure*}

\subsection{The IR softness diagram}

In order to investigate whether the similar $\eta\prime_{IR}$ values found in (U)LIRGs and  the rest of the SFGs can be interpreted
as a similar hardness of the incident radiation,
we compared the observed emission-line ratios in the softness diagram (i.e. separating each emission-line ratio on a different axis of a 2D plot) to allow a better comparison with models.
In the  panels of Figure \ref{eta-sf} we show this diagram in the IR,  representing the emission-line ratios [\neii] $\lambda$ 12.8 $\mu$m/[\neiii] $\lambda$ 15.6 $\mu$m and [\siii] $\lambda$ 18.7 $\mu$m /[\siv] $\lambda$ 10.5 $\mu$m for both SFGs and (U)LIRGs.
We also show in the same figure the mean values of log($\eta\prime_{IR}$) for the subsamples of SFGs (solid black line) and (U)LIRGs (dashed line). 

The data is compared in the panels of Figure \ref{eta-sf} with different sequences of photoionisation models, {with the lines joining points with the same assumed $T_*$. 
Panels a, b, and d show models from 30 kK to 60 kK for WM-Basic \citep{wmbasic} stellar atmospheres and from 80 kK to 120 kK for central stars in PNe \citep{rauch} atmospheres, while panel c shows
models calculated assuming a black body with $T_*$ from 30 kK to 100 kK.} 
As expected, the sequences with higher $T_*$ tend to be in the lower right part of the diagram, corresponding thus to lower values of $\eta\prime_{IR}$.
The points in each sequence of $T_*$ represent different values for log$U$ from -4.0 (upper right in each sequence) to -1.5 (lower left). 
{The effect of considering different metallicities in the models is illustrated in panels a and c  by means of solid lines for 12+log(O/H) = 8.9, while the dashed lines represent a value of 12+log(O/H) = 8.0. 
Panel b of the same figure shows  model sequences calculated for 12+log(O/H) = 8.6 and for different geometries. In this case the solid lines represent models calculated with spherical geometry, while the dashed lines represent models with plane-parallel geometry.
Finally, panel d shows the effect of considering a standard dust-to-gas mass ratio in the models, represented with solid lines, in comparison with models assuming a doubled proportion of dust.}

As   can be seen from   the  panels of this figure, all objects including (U)LIRGs  are well covered by the grid of models with $T_*$ $<$ 60 kK, irrespective of the assumed SED, metallicity, geometry, {or dust-to-gas ratio}. This could be indicative that massive stars are the main contributors to the ionisation of the gas in these objects and that post- asymptotic giant branch (post-AGB) old stars do not have   any preponderance for the ionisation of any of these galaxies. 

In addition, the obtained differences in the sequences when different geometries or dust-to-gas ratios are considered are not significant enough to imply large deviations in the derived $T_*$ when using these sets of lines. Moreover, the inspection of both samples of objects compared to the sequences of models does not seem to shed any light on the role of metallicity, given the very {similar} average O/H values in the samples of (U)LIRGs  compared with the rest of the SFGs.
On the contrary, the slightly lower $\eta\prime_{IR}$ value found for (U)LIRGs can also be due to a different average excitation of the gas, which  is easily appreciated with the different slopes of the sequences of models  compared with the values for $\eta\prime_{IR}$, as lower values of log $U$ can also lead to lower values for $\eta\prime_{IR}$.

Therefore, in order to correctly interpret if the  observed difference of $\eta\prime_{IR}$ for (U)LIRGs is relevant, it is necessary to compare this observational trend with the predictions from large grids of photoionisation models that take into account the differences in metallicity and excitation to provide an absolute scale of $T_*$ in both samples. 

\begin{figure}
   \centering
\includegraphics[width=8cm,clip=]{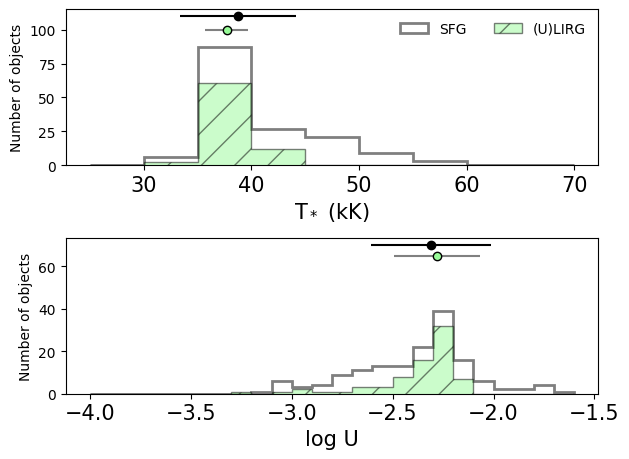}

\caption{Resulting distributions of the solutions found by {\sc HCm-Teff-IR} for $T_*$ (upper panel) and log$U$ (lower panel) when assuming  WM-Basic SEDs and a spherical geometry. The black empty histogram represents SFGs, while the hatched green histogram represents (U)LIRGs. The average values for both distributions are also shown.}

    \label{hist-Teff}%
    \end{figure}

\subsection{Results from {\sc HCm-Teff-IR}}

In order to obtain numerical estimations both for $T_*$ and log $U$, with their corresponding uncertainties, we applied the above-described code {\sc HCm-Teff} for the IR to the compiled sample of SFGs and (U)LIRGs with available IR emission lines. We used whenever possible all lines relevant for our calculation, not just restricting the input for the code to the emission lines involved in the softness diagram shown in Figure \ref{eta-sf}. This implies  considering both [\siii] lines in the IR (at 18.7$\mu$m and 33.5$\mu$m) or the ratios [\oiii]/[\oiv] or [\arii]/[\ariii]. We also introduced as input for the calculations the total oxygen abundances derived using {\sc HCm-IR} and described in Section 2. As also seen in Figure \ref{eta-IR}, a dependence on metallicity exists, so the code can use this value to interpolate the grid of models for each object, using the corresponding uncertainty in the final dispersion reached after the Monte Carlo iteration.

\begin{table}
\begin{center}
\caption{Means and dispersions of the resulting $T_*$ and log $U$ values obtained from {\sc HCm-Teff}  for the subsamples of SFGs and (U)LIRGs   analysed here under different assumptions of SED, geometry (sph stands for spherical and pp for plane-parallel), {or dust-to-gas ratio (a $d/g_{MW}$ value is assumed if   not specified)}. We also list the average metallicity for each set.}

\begin{tabular}{lcc}
\hline
\hline
    &   SFG   &   ULIRG  \\
\hline
N  &  228   &   75  \\
12+log(O/H)  &  8.46 (0.23)  &  8.56 (0.18) \\
log $\eta\prime_{IR}$   &   -0.51 (0.38)  &  -0.57 (0.28)  \\
\hline
  &  \multicolumn{2}{c}{WMB (30-60 kK)} \\
$T_*$ (kK), pp  &  42.3 (3.8)  &   41.1 (1.4) \\
log $U$, pp  &   -2.53 (0.29) &  -2.61 (0.16)   \\
$T_*$ (kK), sph  &  39.9 (4.7)  &  38.0 (2.0)    \\
log $U$, sph  &  -2.37 (0.27)    &  -2.36 (0.21) \\
$T_*$ (kK), sph, 2$\cdot d/g_{MW}$  &  41.8 (4.7)  &  39.8 (1.6)    \\
log $U$, sph, 2$\cdot d/g_{MW}$  &  -2.36 (0.32)    &  -2.42 (0.17) \\
\hline
  &  \multicolumn{2}{c}{WM-Basic (30-60 kK) + Rauch (80-120 kK)} \\ 
$T_*$ (kK), pp  &   53.1 (3.0)  &  51.8 (4.4)  \\
log $U$, pp &     -2.63 (0.27) &   -2.70 (0.15) \\
$T_*$ (kK), sph  &    48.1  (8.8) &  45.6 (3.9) \\
log $U$, sph  &   -2.47 (0.25)  &   -2.45 (0.19)  \\
$T_*$ (kK), sph, 2$\cdot d/g_{MW}$  &    51.4  (9.9) &  49.2 (4.5) \\
log $U$, sph, 2$\cdot d/g_{MW}$  &   -2.48 (0.28)  &   -2.51 (0.15)  \\
\hline
  &  \multicolumn{2}{c}{BB (30-100 kK)} \\ 
$T_*$ (kK), sph  &   51.3 (8.1)  &  51.1 (10.0)  \\
log $U$, sph &  -2.01 (0.52)  &  -2.15 (0.31)  \\
$T_*$ (kK), pp &    52.1 (8.4)  &    51.5 (5.0)  \\
log $U$, pp &   -2.67 (0.27)  &   -2.71 (0.15)  \\
\hline
\label{results}
\end{tabular}
\end{center}
\end{table}

In Figure \ref{hist-Teff} we show the histograms of the resulting $T_*$ and log $U$ obtained using {\sc HCm-Teff} for the samples of SFGs and (U)LIRGs currently analysed here when the code assumes only models with WM-Basic atmospheres and a spherical geometry. The mean of these distributions, along with the 1$\sigma$ dispersions are also listed in Table \ref{results} separately for SFGs and (U)LIRGs. In the same table we also provide the values obtained from the code when other models are assumed. As  can be seen, most of the results predict a lower mean $T_*$ value for (U)LIRGs than for the rest of the SFGs, with a mean offset of around 2 kK, with the only exception of the models assuming black-body SEDs, for which the difference is lower than 1 kK. These average differences are always lower than the corresponding associated dispersions, reinforcing the idea of a similar hardness of the field of radiation in both type of objects.
At the same time, the difference in log $U$ is negligible for models with spherical geometry, while there is a systematic trend to find lower log $U$ values for (U)LIRGs for models with plane-parallel geometry.

The absolute $T_*$ scale strongly depends on the assumed models, as it increases when the models with very high $T_*$ are considered in the calculations. Nevertheless, as seen in Figure \ref{eta-sf}, since the objects in the sample do not lie in regions of the diagram corresponding to these very high $T_*$ values, the inclusion of these sequences in the calculations can artificially increase the scale of the solutions. This scale is also dependent on the geometry, as the sequences corresponding to plane-parallel geometry predict on average lower values for $\eta\prime_{IR}$ than for spherical.
However, it is not expected that the mean $T_*$ difference found between (U)LIRGs and the rest of the SFGs can be due to different geometries in the ionisation of their nebulae, although the influence on geometry  of the strong winds associated with the ionising source should  also be further investigated.

Furthermore, although the average found lower absolute $T_*$ in (U)LIRGs using this procedure looks to be robust as it is  independent of the assumed set in the considered models, the associated errors do not allow us to conclude that it can be real. 
Consequently, the result obtained from {\sc HCm-Teff-IR} using IR lines supports the trend suggested by the similar value, within the errors, found for $\eta\prime_{IR}$ in (U)LIRGs, indicative of a similar incident SED in these objects. 
In any case,  the comparison between the observations and the sequences of models calculated for different values of metallicity and excitation help to discard the possibility of a harder field of radiation in (U)LIRGs considering their average higher metallicity.
On the contrary, a log $U$ tending to be lower and extending over a smaller range in (U)LIRGs is what it seems to contribute to their lower average $\eta\prime_{IR}$.

Our result pointing to a similar $T_*$ in (U)LIRGs  could imply
that the  number of massive stars in these objects is not greater than in the rest of the SFGs, and consequently  
that there are not  significant changes in the slope of the initial mass function (IMF) in the episodes of
star formation that occurred in (U)LIRGs as a consequence of the mergers or interactions processes, 
contrary to some observations indicating a larger number of the most massive stars in these objects \citep{zhang18}.

\begin{figure}
   \centering
\includegraphics[width=8cm,clip=]{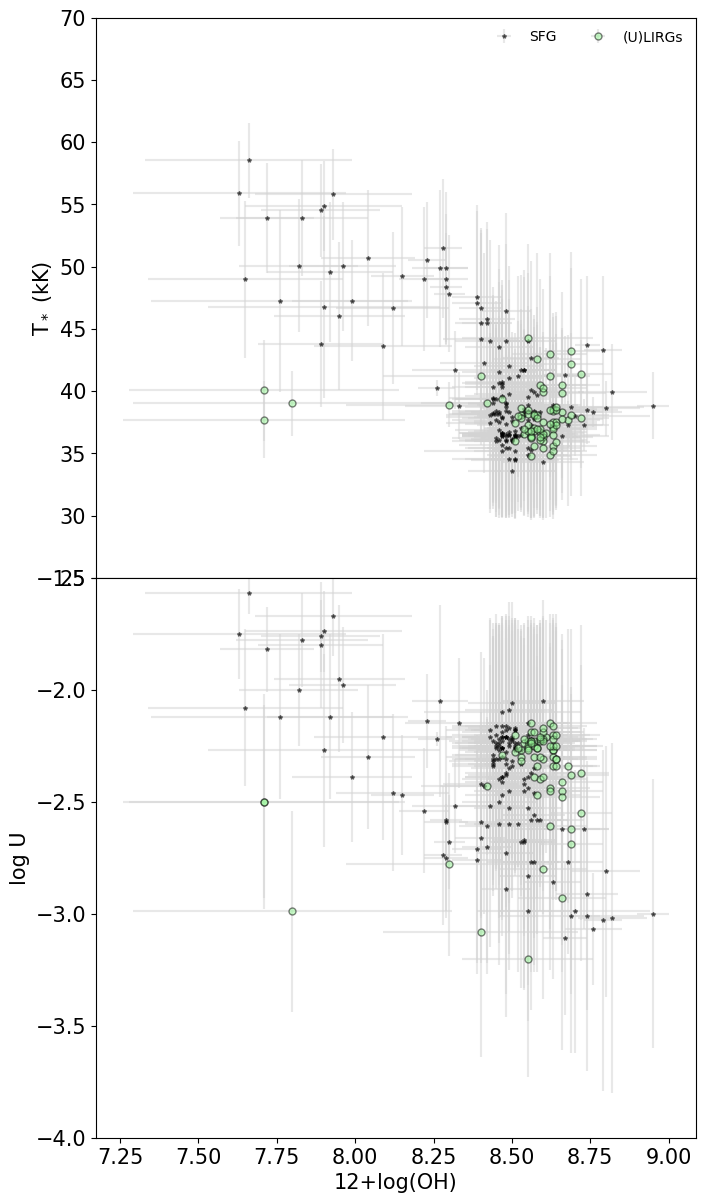}

\caption{Relation between the derived total oxygen abundance with $T_*$ (upper panel) and with log $U$ (lower panel) for our sample of star-forming galaxies (stars) and (U)LIRGs (green circles). The results were obtained using {\sc HCm-Teff} assuming WM-Basic stellar atmospheres with a spherical geometry.}

              \label{OH-Teff-U}%
    \end{figure}

\subsection{Relationship between $T_*$, $U$, and metallicity}

In this subsection we study for our sample of objects the relationship  between the physical conditions derived using {\sc HCm-Teff} from their IR lines, including the excitation of the ionised gas or the hardness of the incident SED, and other properties in the nebulae, like their metal contents.

As described above, the metallicity of the gas can be important to convert the measured values of $\eta\prime_{IR}$ into a quantitative scale for $T_*$. In this way, although the difference between the mean value for $\eta\prime_{IR}$ for (U)LIRGs (i.e. -0.57) and for the rest of the SFGs (-0.51), {as  can also be seen in Figure \ref{eta-OH}}, is not significant, 
this could be incorrectly interpreted as a harder incident SED for (U)LIRGs,
considering the average greater metal content of (U)LIRGs (12+log(O/H) = 8.56), than for SFGs (8.46). 
On the contrary, the Bayesian-like comparison with different grids of models that consider variations for both O/H and log$U$, apparently indicate that $T_*$ is similar, with a slight trend to be lower, in (U)LIRGs. 

On the other hand, the relationship between O/H and $T_*$ appears to be consistent across different studies using the same methodology, based on results obtained from both optical and infrared lines. This relationship indicates that, on average, lower metallicity values correspond to higher $T_*$ values. Consequently, the observed radial gradients of increasing $\eta\prime$ in spiral galaxies \citep{pmv09}, which have been confirmed to be caused by a hardening of the incident radiation field \citep{hcm-teff}, are correlated with gradients of decreasing metallicity as the galactocentric distance increases. This trend also appears to hold when the softness parameter is calculated using IR lines in several spiral galaxies \citep{pmv09}.

We show in the upper and lower panels of Figure \ref{OH-Teff-U} the relationship between the derived $T_*$ and log $U$, respectively, with the total oxygen abundance for the subsamples of (U)LIRGs and for the rest of the SFGs. The Spearman correlation coefficient ($\rho_S$) for this latter subsample of SFGs (-0.61) is consistent with this trend of higher $T_*$ for lower O/H. However, this is not the case for (U)LIRGs, for which almost no correlation is found ($\rho_S$ = -0.07). A similar trend is found for the case of log $U$, which presents a negative correlation with O/H for SFGs ($\rho_S$ = -0.41), while the trend is the opposite for (U)LIRGs ($\rho_S$ = 0.30).

A correlation between the metal content of the gas and its excitation in the sense that the gas could be less excited when metallicity increases has been reported by many authors   based on the optical (e.g. \citealt{pilyugin01}) and on the IR (e.g. \citealt{giveon2002}), but a physical explanation for this relation is not clear.
{Some authors have even proposed that the relation could be sensitive to the assumed SEDs in the models,  that the analysis of the different  emission-line ratios used to trace the excitation could lead to different results (e.g. \citealt{ji22}), or that it is an effect very sensitive to geometry \citep{kewley19}.
On the contrary,  our results seem to indicate that this relation could  more probably be  caused by the more robust relation} between $T_*$ and O/H, and the fact that most of the ratios of intensity of low-to-high excitation emission lines traditionally used to estimate $U$ also depend on $T_*$, as   can be easily confirmed by a visual inspection of the softness diagram. The assumption of an empirical relation between O/H and $U$ made  by some recipes to derive the oxygen abundance from the relative intensities of certain collisionally excited lines (e.g. {\sc HCm}, \citealt{hcm14}) is then mostly motivated by the use of a fixed ionising SED.
On the other hand, the relationship between $T_*$ and O/H could be physically well motivated by the simultaneous presence of heavy elements in the ionised gas and the massive star atmospheres or in their ejected stellar winds that are blanketed and whose UV part of the SED is attenuated, producing an average lower effective temperature. 

Nonetheless, no correlation at all between $T_*$ and O/H is observed in our sample of (U)LIRGs, being especially revealing that   for the subsample of four objects with very low metallicity (i.e. 12+log(O/H) $<$ 8.2, and in three objects even lower than 8.0)
we do not obtain a higher value of $T_*$  compared to the rest of the sample of metal-rich (U)LIRGs. According to \cite{pd23a}, these metal-poor ULIRGs are experiencing a "deep diving" phase under the mass-metallicity relation (MZR) (i.e. a rapid decrease in their metal content due to the dilution of the gas after a merger or interaction event).

\subsection{The role of dust in $T_*$ derivation}

The lack of correlation of $T_*$ with O/H in (U)LIRGs and the similar or even lower  average $T_*$ in these objects  compared with other SFGs could   thus be an indications that another mechanism than a dependence on metallicity is responsible for the found differences.

Among other causes, the presence of large amounts of heated dust grains mixed with the surrounding gas, which is the main cause of the very high IR luminosities observed in (U)LIRGs \citep{sm96}, could be behind the attenuation of the ionising high-energy flux of the innermost stars if the dust is well mixed inside the galaxy. It is known that dust attenuation can be related with  star formation rate (SFR) for high stellar-mass galaxies \citep{zahid13}, and the SFRs measured in (U)LIRGs are very high   as a consequence of  the large amount of available gas and the intensity of the gas interchange between the galaxies taking place in the merger or interaction process (e.g. \citealt{piqueras16}).
In the case of our subsamples, the mean measured SFRs confirm this very large
difference (i.e. around 100 M$_{\odot}$/yr for the (U)LIRGs, and 3 M$_{\odot}$/yr for the rest of the SFGs, according to \citealt{pd23a}).  
 Our result seems to validate this hypothesis as, contrary to most SFGs,  no large variations in $T_*$ are found in our sample of (U)LIRGs and no direct relation between $T_*$ and O/H is found in our sample.

{Given that it seems clear that (U)LIRGs are dust-bound  objects (e.g. \citealt{voit92, abel03}), although no clear results are found regarding a higher dust-to-gas mass ratio in these objects \cite{Herrero-Illana2019}},  
it is pertinent to wonder if this amount of extra dust mixed with the gas, and deeply affecting both the incident field of radiation and the optical lines, can affect the derived metallicity or $T_*$ based on photoionisation codes.
However, on the contrary, the use of IR lines  prevents   this effect as their fluxes are much less attenuated than in the case of optical lines, providing a more accurate determination of the metallicity of the gas even in regions behind a large optical depth. In addition, the consistent position of this sample of (U)LIRGs  compared with other SFGs of their same stellar mass in the MZR; \citealt{pd23a}) seems to indicate that no large variations in the final derived O/H is found when this extra innermost amount of dust is considered in the models.

{In the case of $T_*$, we computed additional models considering a $d/g$ mass ratio twice the standard  galactic value considered in the rest of the models to verify the role of this parameter in our results. As can be seen in panel d of Figure \ref{eta-sf}, no large difference is seen in the softness diagram. Regarding our results with the code when we use this grid of models, as can also be seen in Table \ref{results}, there is a trend to find slightly higher $T_*$ values ($\sim$ 2 kK) when these models with higher $d/g$ are considered. Moreover, although this difference is never larger than the obtained errors,  it could be even behind the lower mean $T_*$ values obtained for (U)LIRGs when the same $d/g$ is assumed, reinforcing thus the conclusion of a similar hardness of the field of radiation in both samples, at least as seen from the IR lines emitted by the gas-phase in each galaxy.}

\section{Summary and conclusions}

In this work we explored the softness diagram based on mid-IR emission lines, less affected by dust obscuration and with a lower dependence on electron temperature than optical emission lines. In addition, the exploitation of this spectral range for nearby star-forming galaxies allows  the  use of  emission lines at a higher ionisation stage, such as [\siv] at $\lambda$ 10.5 $\mu$m,
which in turn means that this diagram has a lower dependence on diffuse emission.

The IR softness parameter, $\eta\prime_{IR}$, based on the relative intensities of the IR lines of [\neii], [\neiii], [\siii], and [\siv], traces the hardness of the incident radiation field, but it also shows a certain dependence on excitation and, although to a lesser extent than its equivalent form in the optical, on metallicity. For this reason we developed the code \textsc{HCm-Teff-IR}, which performs a Bayesian-like comparison between observed emission-line ratios, involved in $\eta\prime_{IR}$, such as  [\neii]/[\neiii] and [\siii]/[\siv], but also including other IR emission-line ratios such as [\ariii]/[\ariv] and [\oiii]/[\oiv],
 with a large grid of photoionisation models calculated assuming a central single star to provide an estimation of the equivalent effective temperature ($T_*$) and the ionisation parameter (log $U$).

We applied this code to a sample of collected SFGs and (U)LIRGs with available IR observations. The mean observed $\eta\prime_{IR}$ is similar within the errors
in (U)LIRGs than in the rest of the SFGs, which could be interpreted as  the field of radiation  also being similar.
This trend is confirmed by  the analysis of the compiled subsamples with \textsc{HCm-Teff-IR} code, that  reveals that the obtained mean $T_*$ is similar within the errors for this type of objects  compared with the rest of the SFGs.
In both cases we obtain values of $T_*$ compatible with the ionisation from massive stars and, contrary to the equivalent diagnostics based on optical emission lines, the contribution from
other harder ionisation sources (e.g. post-AGB stars) is not required.

Regarding the observed relation between the obtained $T_*$ and metallicity, contrary to the rest of the SFGs for which higher $T_*$ values are found on average
for lower oxygen abundances,
$T_*$ does not correlate with metallicity in (U)LIRGs. This result, combined with the fact that $T_*$ is similar   or  even slightly lower in (U)LIRGs, contrary to expected given the very high SFRs measured in these objects,
can be interpreted in terms of a much higher small heated dust grain proportion in their high SFR environments, producing a blanketing of the higher energetic range of the massive stellar populations, {although this is not evidenced from the assumption of a higher dust-to-gas mass ratio in the gas-phase considered in the models, which predict a very slight increase within the errors  in the mean $T_*$.}

These results could largely be reinforced with the addition of larger samples in this spectral range, combining spatially resolved observations that help to disentangle between young and  old stellar populations, and the regions with a larger dust presence.
An equivalent analysis for sources at higher redshift would depend on the availability of data at a longer wavelength, which unfortunately at the present moment is not covered by any facility, but at least the inclusion in our code of Ar lines in combination with Ne lines, would   extend the analysis up to $z \sim$ 1.5 using JWST.

\begin{acknowledgements}
This work has been partly funded by projects Estallidos7 PID2019-107408GB-C44 (Spanish Ministerio de Ciencia e Innovacion), and the Junta de Andaluc\'\i a for grant P18-FR-2664.
We also acknowledge financial support from the State Agency for Research of the Spanish MCIU through the "Center of Excellence Severo Ochoa" award to the Instituto de Astrof\'\i sica de Andaluc\'\i a  (SEV-2017-0709).
"JAFO acknowledges financial support by the Spanish Ministry of Science and Innovation (MCIN/AEI/10.13039/501100011033) and ``ERDF A way of making Europe'' though the grant PID2021-124918NB-C44; MCIN and the European Union -- NextGenerationEU through the Recovery and Resilience Facility project ICTS-MRR-2021-03-CEFCA.
EPM also acknowledges the assistance from his guide dog Rocko without whose daily help this work would have been much more difficult.

\end{acknowledgements}

%
   \bibliographystyle{aa} 
   \bibliography{refs.bib} 
%

\end{document}